\documentclass[10pt,conference]{IEEEtran} 
\IEEEoverridecommandlockouts

\def\BibTeX{{\rm B\kern-.05em{\sc i\kern-.025em b}\kern-.08em
    T\kern-.1667em\lower.7ex\hbox{E}\kern-.125emX}}

\usepackage{algorithmic}
\usepackage{amsfonts}
\usepackage{amsfonts}
\usepackage[english]{babel}
\usepackage{comment}
\usepackage{courier}
\usepackage{csquotes}
\usepackage{epigraph}
\usepackage{float}
\usepackage[T1]{fontenc}
\usepackage[keeplastbox]{flushend}
\usepackage{graphicx}
\usepackage[utf8x]{inputenc}
\usepackage{layout}
\usepackage{listings}
\usepackage{mdframed}
\usepackage{paralist}
\usepackage{pifont}
\usepackage{subfigure}
\usepackage{textcomp}
\usepackage{tikz}
\usepackage{url}
\usepackage{xcolor}

\usepackage{hyperref}
\AtBeginDocument{%

}

\usepackage{ifdraft}
\usepackage{ifthen}

\newboolean{debug}
\setboolean{debug}{false}

\ifthenelse{\boolean{debug}}{%
  \usepackage{showframe}                 
  \usepackage{todonotes}                 
  \newcommand{\xynote}[2]{\todo[inline]{#1: #2}}
  
  \usepackage{hyperref}                  
  \pagestyle{plain}                      
  \usepackage{setspace}                  

}{%
  \usepackage[disable]{todonotes}        
  \newcommand{\xynote}[2]{}
  
  \hypersetup{hidelinks}                 
  \pagestyle{empty}                      
}

\newcommand{\journal}[1]|{}               

\usepackage{xspace}
\usepackage{ifdraft}

\newcommand{\mc}{Monti\-Core\xspace}

\newcommand\blfootnote[1]{%
  \begingroup
  \renewcommand\thefootnote{}\footnote{#1}%
  \addtocounter{footnote}{-1}%
  \endgroup
}

\newcommand*{\ie}{\textit{i.e.,}\@\xspace}
\newcommand*{\eg}{\textit{e.g.,}\@\xspace}
\newcommand*{\cf}{\textit{cf.}\@\xspace}

\makeatletter
\newcommand*{\etc}{%
  \@ifnextchar{.}%
  {\textit{etc}}%
  {\textit{etc.}\@\xspace}%
}
\makeatother

\definecolor{se-green}{RGB}{0,128,0}
\definecolor{se-blue} {RGB}{0,0,204}

\newcommand{\code}[1]{\texttt{#1}}
\newcommand{\cw}[1]{\code{#1}}

\newcounter{requirement}[section]

\newenvironment{requirement}[1]{%
  \refstepcounter{requirement}%
  \renewcommand*{\theenumi}     {\textit{#1\arabic{enumi}}}%
  \renewcommand*{\theenumii}    {\textit{.\arabic{enumii}}}%
  \renewcommand*{\theenumiii}   {\textit{.\arabic{enumiii}}}%
  \renewcommand*{\labelenumi}   {\textbf{#1\arabic{enumi}}}%
  \renewcommand*{\labelenumii}  {\textbf{#1\arabic{enumi}.\arabic{enumii}}}%
  \renewcommand*{\labelenumiii} {\textbf{#1\arabic{enumi}.\arabic{enumii}.\arabic{enumiii}}}%
  \enumerate
}{%
  \endenumerate
}

\usepackage{listings}
\newenvironment{ownfigure}[0]%
{\begin{figure}[!ht]}
{\end{figure}}

\usepackage{color}
\definecolor{DarkRed}{rgb}{0.75,0,0}
\definecolor{Lightgreen}{rgb}{0.588,1.0,0.588}
\definecolor{DarkGreen}{rgb}{0,0.5,0}
    
\lstdefinelanguage{MontiArc}[]{Java}{
  morekeywords={component, port, in, out, inv, package, import, connect, autoconnect}
}

\lstdefinelanguage{myJava}[]{Java}{
  commentstyle=\color{DarkGreen}\itshape 
}

\lstdefinelanguage{MontiArcAutomaton}[]{Java}{
  morekeywords={component, port, in, out, inv, package, import, connect,
  autoconnect, automaton, state, ocl, java, initial, final,
  noCompletion, chaosCompletion, var, mode, activate, transitions,
  modetransitions}, commentstyle=\color{DarkGreen}\itshape }

\lstdefinelanguage{mcSC}[]{}{
	morekeywords={package, statechart, initial, state, final}
}

\lstdefinelanguage{mcTag}[]{}{
	morekeywords={conforms, to, tags, tag, for, with, within, in, out}
}

\lstdefinelanguage{mcSchema}[]{}{
	morekeywords={tagschema, tagtype, for, in, out}
}

\lstdefinelanguage{MCConfig} { 
    morekeywords={config, Require, Model} 
}

\lstdefinelanguage{Manifest} { 
    morekeywords={Manifest, Bundle, ManifestVersion, Name, SymbolicName,
      Version, Require
    } 
}

\lstdefinelanguage{mcGrammar}[]{}{
  morekeywords={
    grammar, package, path, parser, lexer, nows, noslcomments, nomlcomments, 
    noident, nostring, noanything, nocharvocabulary, dotident, identrule,
    xmlcomments, hashcomments, texcomments, freemarkercomments, concept, 
    globalnaming, define, usage, options, true, false, protected, ident, 
    compilationunit, extends, implements, interface
  }
}

\lstdefinelanguage{mcLng}[]{}{
  morekeywords={
    dsltool, language, package, path, parser, root, parsingworkflow, 
    rootfactory, lexer, nows, noslcomments, nomlcomments, noident, nostring,
    dotident, concept, globalnaming, define, usage, options, true, false, 
    protected, ident
  }
}

\lstdefinelanguage{mcManifest}[]{}{
  morekeywords={
    bundle, Bundle, Name, SymbolicName, true, false, Main, Class, 
    Version, Activator, Localization, Require, 
    Exclude, Eclipse, LazyStart, Vendor, Export, Package, 
    ClassPath
  }
}

\lstdefinelanguage{Alloy}[]{Java}{
commentstyle=\color{DarkGreen}\itshape,
  morekeywords={abstract,sig,->,fact,pred,fun,run,for,iff,
  not,no,one,all,some,lone,\#,set,in,and,or,but,exactly,none,univ,Int,assert,check},
  otherkeywords = {[2]????},
    morekeywords = {[2]????},
    keywordstyle={[2]\color{blue}},
    otherkeywords = {[3]????,<,<->,->, &, |, =, !=, !,<:,~},
    morekeywords = {[3]????,<,<->,->, &, |, =, !=, !,<:,~},
    keywordstyle={[3]\color{blue}}
  }

\lstdefinelanguage{mccd}[]{Java}{
  morekeywords={classdiagram,abstract,<<singleton>>,class,int,String,
  association,composition,extends}
}

\lstdefinelanguage{FreeMarker}[]{}{
  keywordsprefix={\#},
  keywords={in},
  commentstyle=\color{DarkGreen}\itshape }

\lstdefinelanguage{Mona}[]{}{
  morekeywords={ex0,all0,ex1,all1,ex2,all2,var0,var1,var2,pred,in,notin,include,union,inter,empty,assert},
  morecomment=[l]{\#},
  commentstyle=\color{DarkGreen}\itshape,
  otherkeywords = {[2]????,next,boolean,init,case,esac},
  morekeywords = {[2]????,next,boolean,init,case,esac},
  otherkeywords = {[3]????,<,<=>,=>, &, |, =, !=, !},
  morekeywords = {[3]????,<,<=>,=>, &, |, =, !=, !},
}

\lstdefinelanguage{myPython}[]{Python}{
  morekeywords={assert},
  morecomment=[l]{\#},
  commentstyle=\color{DarkGreen}\itshape,
}

\lstdefinelanguage{GeneratorConfiguration}[]{Java} {
  morekeywords={
    template, 
    generator, 
    ast, 
    runtime},
}

\lstdefinelanguage{ApplicationConfiguration}[]{Java} {
  morekeywords={
    application,
    behaviors,
    bindings,
    classdiagrams,
    components,
    factories,
    generators,
    map,
    to},
}

\lstdefinelanguage{Isabelle}[]{} {
    morekeywords={
        datatype,
        typedef},
}

\usepackage[nomain,acronym]{glossaries}

\newcommand{\singular}[1]{\gls{#1}\xspace}
\newcommand{\plural}[1]{\glspl{#1}\xspace}
\newcommand{\abkuerzung}[2]{\newacronym{#1}{#1}{#2}}

\abkuerzung{AADL}{Architecture Analysis \& Design Language}

\abkuerzung{CBL}{Case-Base Language}
\newcommand{\cbl}{\singular{CBL}}

\abkuerzung{CBR}{Case-Based Reasoning}
\newcommand{\cbr}{\singular{CBR}}

\abkuerzung{CD}{class diagram}
\newcommand{\cds}{\plural{CD}}

\abkuerzung{CPPS}{Cyber-Physical Production System}
\newcommand{\cpps}{\singular{CPPS}}
\newcommand{\cppss}{\plural{CPPS}}

\abkuerzung{CSL}{Case Similarity Language}
\newcommand{\csl}{\singular{CSL}}

\abkuerzung{DSL}{Domain-Specific Language}
\newcommand{\dsl}{\singular{DSL}}
\newcommand{\dsls}{\plural{DSL}}

\abkuerzung{DS}{Digital Shadow}

\abkuerzung{DT}{Digital Twin}
\newcommand{\dt}{\singular{DT}}
\newcommand{\dts}{\plural{DT}}

\newcommand{\ind}{Industry 4.0\xspace}

\abkuerzung{MA}{MontiArc}

\abkuerzung{MDD}{Model-Driven Development}

\abkuerzung{MDSE}{Model-Driven Systems Engineering}

\abkuerzung{OCL}{Object Constraint Language}

\abkuerzung{PDDL}{Planning Domain Definition Language}
\newcommand{\pddl}{\singular{PDDL}}

\abkuerzung{SLE}{Software Language Engineering}

\abkuerzung{SysML}{Systems Modeling Language}

\abkuerzung{UML}{Unified Modeling Language}

\usepackage{booktabs}

\begin{document}

\title{Self-Adaptive Manufacturing with Digital Twins}


%


\author{
\IEEEauthorblockN{
Tim Bolender\IEEEauthorrefmark{1},
Gereon Bürvenich\IEEEauthorrefmark{1},
Manuela Dalibor\IEEEauthorrefmark{1},
Bernhard Rumpe\IEEEauthorrefmark{1}, 
Andreas Wortmann\IEEEauthorrefmark{1}\IEEEauthorrefmark{2}\\
}
\\
\IEEEauthorblockA{
\IEEEauthorrefmark{1} Software Engineering,
RWTH Aachen University, Aachen, Germany, \url{www.se-rwth.de}
}
\IEEEauthorblockA{
\IEEEauthorrefmark{2} Institute for Control Engineering of Machine Tools and Manufacturing Units\\
University of Stuttgart, Stuttgart, Germany, \url{www.isw.uni-stuttgart.de}
}
}






\maketitle 
\begin{abstract}
	%
	Digital Twins are part of the vision of Industry 4.0 to represent,
	control, predict, and optimize the behavior of \cppss.
	%
	%
	These \cppss are long-living complex systems deployed to and configured for
	diverse environments.
	Due to specific deployment, configuration, wear and tear, or other
	environmental effects, their behavior might diverge from the intended behavior over time.
	Properly adapting the configuration of \cppss then relies on the expertise of
	human operators.
	%
	%
	Digital Twins (DTs) that reify this expertise and learn from it to address
	unforeseen challenges can significantly facilitate
	self-adaptive manufacturing where experience is very specific and, hence,
	insufficient to employ deep learning techniques.
	%
	%
	We leverage the explicit modeling of domain expertise through
	case-based reasoning to improve the capabilities of Digital Twins for
	adapting to such situations.
	%
	%
	To this effect, we present a modeling framework for self-adaptive
	manufacturing that supports modeling domain-specific cases, describing rules
	for case similarity and case-based reasoning within a modular Digital Twin.
	%
	%
	Automatically configuring Digital Twins based on explicitly modeled domain
	expertise can improve manufacturing times, reduce wastage, and, ultimately,
	contribute to better sustainable manufacturing.
\end{abstract}











\section{Introduction}
\label{sec:Introduction}

\ind, the fourth industrial revolution, focuses on integrating Cyber-Physical Production Systems (CPPSs), their
processes, and stakeholders to optimize the complete value-added chain to ultimately
save time, cost, and reduce resource consumption~\cite{WBCW20}.
These \cppss are long-living complex systems deployed to and configured for
diverse environments.
Due to specific deployment, configuration, wear and tear, or other
environmental effects their behavior as-operated can diverge from its
behavior as-designed over time.
Successfully using the \cppss demands the expertise of human operators
to mitigate these effects.
In such cases, experienced operators employ significant manual efforts to
configure the \cppss before starting production.
Making their expertise machine-processable can facilitate their self-adaptive
operations.
%
\blfootnote{Funded by the Deutsche Forschungsgemeinschaft (DFG, German Research 
	Foundation) under Germany's Excellence Strategy - EXC 2023 Internet of 
	Production - 390621612.}

One vision for implementing \ind are so-called \dts~\cite{tao2018digital},
which are digital duplicates of \cpps that represent, control, and predict the
behavior of their physical counterparts.
Where \dts control \cppss, they need to have knowledge about the \cpps and
its operations.
Consequently, they are connected to the \cpps and lend themselves for
automatically adapting it to changing challenges.
For their adaptation of the overall manufacturing system consisting of \dt
and \cpps, the \dts need to be enabled to sense changes in the \cpps's
behavior, reason over reified domain expertise, and make changes to the
\cpps accordingly. 

\cbr~\cite{AP94,Rie13} is an AI planning technique in which highly
specific, and hence, sparse domain expertise is reified in cases.
These describe undesired situations a system should react to together with
suitable reactions to capture the expertise of \cpps operators.
Based on this domain expertise, the \dt of a \cpps can detect undesired
situations, find matching or similar cases, and adjust the \cpps according
to their reactions to produce a desired system state again.

This enables integrating highly domain-specific expertise (that can hardly
be foreseen by the \cppss' developers) in brownfield settings where the
long-living \cppss are already in place as well as in greenfield settings,
where the \cpps and its \dt are developed together.

To leverage \cbr over domain expertise into self-adaptive manufacturing, we
devised a modeling framework comprising multiple interrelated modeling
languages and integrate it into our model-driven architecture for 
\dts~\cite{BDH+20,KMR+20,dalibor2020towards}.
The contributions of this paper are
\begin{itemize}
	\item extensible modeling languages to capture domain expertise in the form of cases 
	and to describe the similarity between them, and 
	\item a modular architecture for integrating \cbr into \dts and supporting all activities 
	related to identifying, applying, and learning cases.
\end{itemize}

In the following, \autoref{sec:Context} illustrates the challenges of
incorporating domain expertise into manufacturing on the example of
injection molding.
\autoref{sec:Preliminaries} then introduces preliminaries.
Afterwards, \autoref{sec:CBR} presents our \cbr modeling languages and
\autoref{sec:Framework} presents our realization of \cbr within \dts.
\autoref{sec:Example} illustrates our method's application to the
configuration of an injection molding process.
\autoref{sec:Discussion} discusses observations and
\autoref{sec:RelatedWork} highlights related research.
\autoref{sec:Conclusion} concludes.

\section{Context}
\label{sec:Context}


Injection molding is a popular form of batch processing for the mass
production of 3D plastic parts that is performed daily billionfold around
the world to manufacture identical parts repeatedly in high quality.
Injection molding itself is highly automated but requires a configuration
that often has to be determined and adjusted manually due to changing \cpps
properties, materials, or environmental characteristics.

\begin{figure}[t] \centering
	\includegraphics[width=\columnwidth]{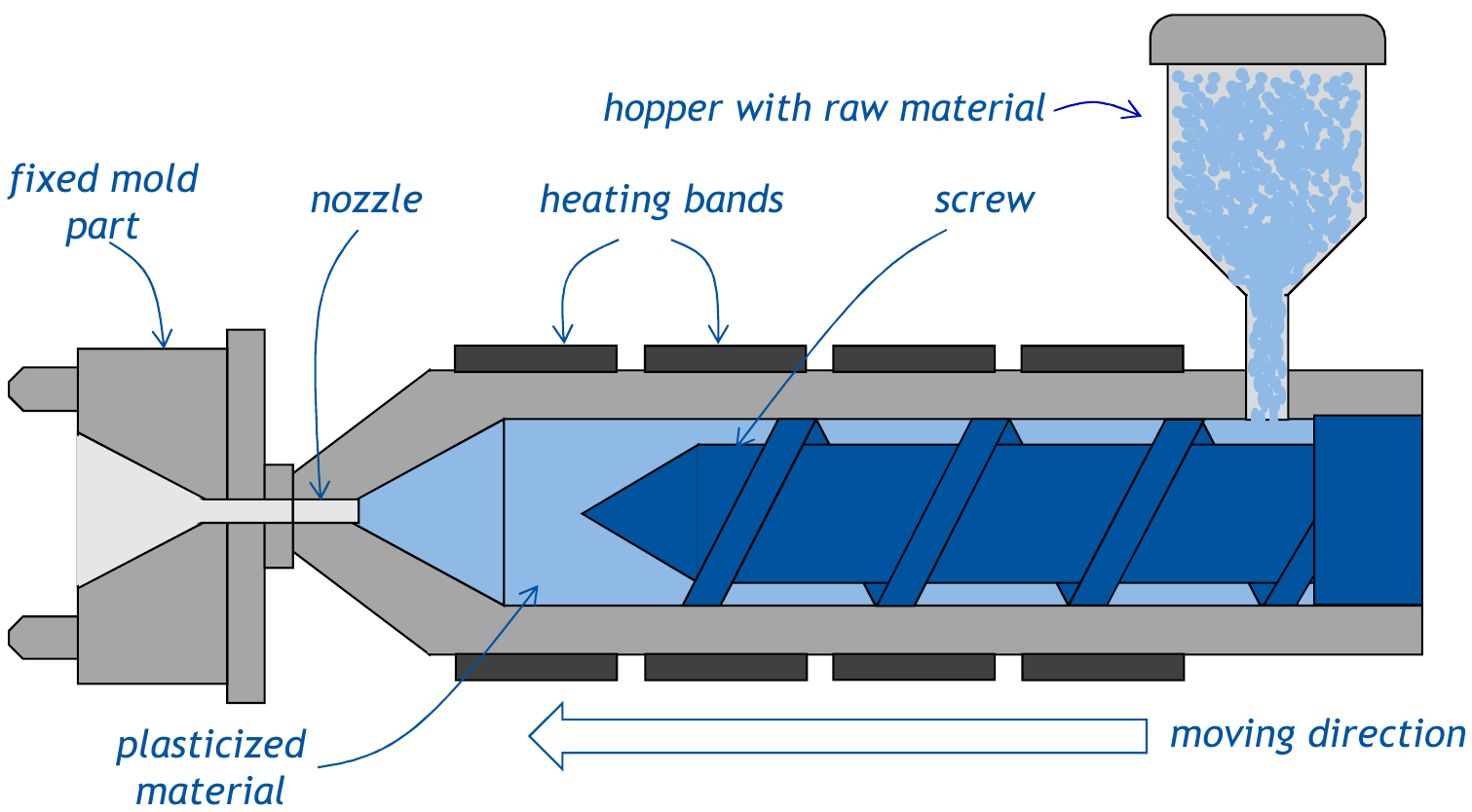}
	\caption{Detailed view of the plasticizing unit of an injection molding
		machine.}
	\label{fig:injection_molding_machine}
\end{figure}

\autoref{fig:injection_molding_machine} illustrates the typical components of 
an injection molding machine.
It consists of a hopper to insert the material into the
plasticizing unit.
The plasticizing unit heats the material until it melts and carries it to the 
front through a screw.
Heating bands support the melting.
Next, the material is injected into the mold, which is the 3D negative of the 
part to be produced.
The material solidifies inside the mold and forms the desired shape.

Overall, the production process consists of four phases:
\begin{inparaenum}[(1)]
	\item Dosing: The material is fed into the cylinder.
	Rotation of the plasticizing unit screw conveys the material to the 
	nozzle.
	Through the heating and the movement friction, the material plasticizes.
	\item Injecting: The screw moves towards the nozzle and injects the 
	material into the mold.
	The movement speed and the viscosity of the material determine the speed of 
	the injection.
	Parameters of this phase are temperature, volume, time, speed, and mold characteristics.
	\item Holding: The screw slows down, and the clamping unit applies pressure 
	to the mold.
	Thereby, the material fills the last parts of the mold.
	The characterizing parameters during the holding phase are the time and 
	pressure.
	\item Cooling: Before ejecting the finished part, the material has to 
	solidify by cooling off.
	The correct time and temperature of the mold prevent defects such as 
	warpage.
\end{inparaenum}

The great variety of environmental and \cpps influences, as well as of process
parameters and their impact on the process and part quality complicates
finding optimal \cpps configurations.
Deviations from the predictions of simulations are common, especially due to
wear and tear.
Consequently, only experienced operators can configure the \cpps properly by
applying domain expertise learned during their career.
\dts can help to overcome these difficulties by reifying the operators'
domain expertise and automatically controlling the \cpps.
We, therefore, identify the following four requirements for incorporating
domain expertise via \dts in batch processing:

\begin{requirement}{R}
	
	\item \textbf{Comprehensibility}: Operator expertise must be reified in 
	means that are comprehensible by domain
	experts and support non-clear cohesion as well as the integration of
	empirical knowledge.
	
	\item \textbf{Automatability}: The \dt has to monitor the situation of
	the underlying \cpps permanently and, when encountering undesired states,
	has to automatically adapt the \cpps without further interaction based on
	the reified domain expertise.
	
	\item \textbf{Adjustability}: The \dt and its domain knowledge have to
	be adjustable to different contexts, deployments, configurations without
	in-depth software engineering expertise.
	
	
	
	\item \textbf{Self-Explainability}: Extracting knowledge is difficult in a
	domain with unclear coherence, yet the \cppss' decisions should be
	comprehensible by domain experts.
	To support self-explainability, the \dt needs to support the creation of
	empirical-based analytical knowledge.
	
\end{requirement}

We conceived extensible modeling languages to fulfill \mbox{R1} and
\mbox{R2}, a modular \dt architecture supporting \mbox{R3}, and a case
synthesis realizing \mbox{R4}.
In the following, we present these modeling languages, an implementation of
our \dt architecture for \cbr, and a system for reifying, applying, and
producing domain expertise through \cbr.
\section{Preliminaries}
\label{sec:Preliminaries}

In our approach to self-adaptive manufacturing, we employ \cbr, AI action
planning, and software language engineering.
This section introduces these preliminaries. 

~\\\noindent\textit{Case-Based Reasoning and Planning}
\vspace{.2em}

\noindent A \dt that controls a \cpps encounters situations that are not
anticipated during specification and thus should adapt to new conditions
autonomously.
\cbr~\cite{AP94} is a problem-solving paradigm that utilizes knowledge about
previously encountered situations and reuses their solutions.
Consequently, a case consists of a situation description (a condition over
available data sources), its solution, and additional information about how
the solution was derived.
The \cbr cycle is divided into four phases:
\begin{inparaenum}[(1)]
	\item Retrieve the case most similar to the current situation.
	\item Reuse the solution of the most similar case.
	\item Revise that solution if the case differs too much from the current
	situation.
	\item Retain the revised case in the knowledge base.
\end{inparaenum}
Hence, an essential part of the \cbr cycle is the identification of similar
cases.
If the case's condition can reference multiple heterogeneous attributes, a
generally useful similarity cannot be specified; instead, this consideration
is highly domain-specific.
To support engineers and domain experts in specifying similarity measures,
these often are broken down according to the \emph{local-global-principle}:
A distinct local measure defines the similarity for each individual
attribute referenced in a case condition.
A global similarity measure then enables computing the similarity for the
whole case by using, \eg the weighted average of all the local similarities.

\cbr, hence, is limited to applying existing cases and learning deviations
of cases.
It cannot, generally, produce new solutions to completely unforeseen
challenges.
General automated planning and scheduling supports creating new solutions
(plans) to unforeseen challenges, if the necessary primitives (types,
actions) are provided.
In our architecture, we leverage AI planning based on the 
\pddl~\cite{fox2003pddl2}
as a fallback mechanism when \cbr fails.
\pddl is a language for representation and exchange of planning domain
models comprising types, constants, and actions with preconditions and
postconditions.
A \pddl problem description is an instantiation of model elements and
formulates a goal that describes which situation the \dt shall achieve.
A planning system, such as MetricFF~\cite{hoffmann2003metric} processes
domain models and problem descriptions and derives a sequence of actions, a
plan, that leads from the initial situation to the goal~\cite{HN11}.
Thus, \pddl can be employed as a fallback if \cbr cannot find a similar case to address undesired situations~\cite{ABH+17}.

~\\\noindent\textit{MontiCore Language Workbench}
\vspace{.2em}

\noindent Our method presented in the following relies on modeling
languages~\cite{HRW18} to describe cases, case similarities, \dt
architecture components, and its ties to \cpps, model transformations, and
code generation.
All of these exist in the technological space of the MontiCore~\cite{HR17}
language workbench~\cite{erdweg2013state} for the efficient engineering of
modular, textual modeling languages.
These modeling languages comprise context-free grammars, Java-based
well-formedness rules, model-to-model transformations, and FreeMarker-based
code generators~\cite{HR17}.
From grammars, MontiCore derives an extensible infrastructure to parse,
check, and transform models of the languages defined by the grammar.
MontiCore comes with a multitude of reusable modeling language modules ranging
from expressions and statements of various complexities, to UML fragments,
software architectures, and more.

A particular kind of languages available in the technological space of
MontiCore are domain-specific tagging languages~\cite{Concept_Tagging15}
that support extending models of a base language with additional information
without polluting these.
To this effect, their infrastructure (grammar, well-formedness rules) is
derived from a base language to enable enriching models of that language
with information, \eg about platform-specific details of their use, without
polluting them.
As the tag model is separate from the base model, models of the base
language are unaware of being tagged and can thus be reused in different
contexts.

~\\\noindent\textit{Digital Twin Architecture}
\vspace{.2em}

\noindent MontiArc~\cite{BHH+17,BKRW17a} is an architecture description language for specifying 
reusable components within a software architecture and their connections 
through typed, directed ports. MontiArc comes with a Java code generator that 
generates Java classes conforming to the specified components, methods to 
access port values, and a mechanism to inject handwritten behavior 
specifications. 
In previous work, we built a \dt with MontiArc, that enables automatic 
experiment execution on injection molding machines~\cite{BDH+20} and also 
provides a cockpit for visualizing the current state of the 
machine~\cite{DMR+20}. 
\begin{figure}[htb] \centering
	\includegraphics[width=\columnwidth]{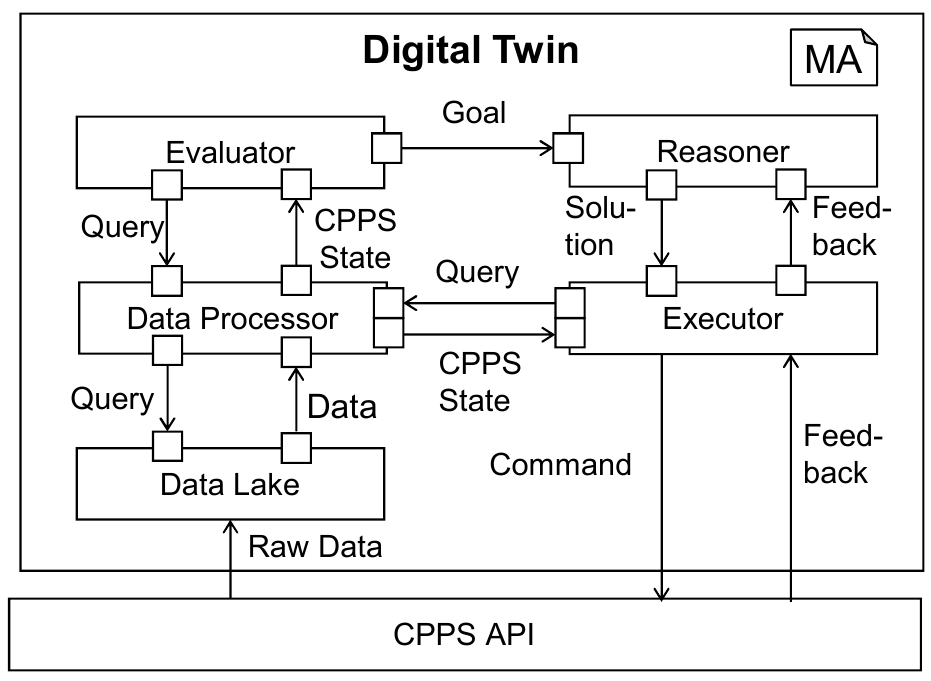}
	\caption{
		Reusable Digital Twin Architecture modeled with MontiArc.
	}
	\label{fig:MD_DT_Architecture}
\end{figure}
We define a Digital Twin (DT) of a system as a 
set of models of the system, a set of contextual data traces, and a set of 
services to use the data and models purposefully with respect to the original 
system~\cite{BDH+20}.
In our notion, a \dt is a software system representing a physical counterpart 
and encapsulating domain knowledge in the form of models that characterizes 
this physical counterpart. Further, it contains data about the physical 
counterpart and services to collect more data or interact with this 
counterpart. 
We built a \dt (\cf
\autoref{fig:MD_DT_Architecture}) with MontiArc that provides the following components:
\begin{itemize}
	\item Data Lake: Encapsulates multiple databases that store data about the 
	physical system, its context, and data produced by the \dt
	\item Data Processor:  Accesses the Data Lake and collects relevant data 
	for the \dt
	\item Evaluator:  Supervises the physical system's state and triggers the 
	reasoner if a malfunctioning is detected
	\item Reasoner:  Based on data about the physical system and models 
	describing its intended behavior, finds a solution to return to the 
	intended 
	state
	\item Executor: Accesses the physical system via OPC UA~\cite{LM06} and 
	ensures that the solution 
	provided by the Reasoner is executed on it.
\end{itemize}
While previous reasoner implementations offered a way to organize experiments, 
we will exchange this Reasoner with a new reasoner that performs \cbr. 
\section{Modeling Languages for Case-Based Reasoning}
\label{sec:CBR}

We present a modeling framework that supports the cycle of \cbr and
supports the creation, storage, retrieval, and comparison of cases via the
case base.
Since the \dt architecture and connectivity to the \cpps are provided by the 
\dt 
framework, domain experts only need to provide the essential domain knowledge, 
defining known
experiences as cases and specifying case similarity measures to create a new 
\dt for a \cpps.
We utilize UML/P \cds \cite{Rum17} and introduce further modeling languages
to support the description of the domain knowledge.
The integration of these models supplements the framework configuration and
incorporates the domain knowledge into the workflow.
Additionally, the framework enables defining \pddl-based fallback strategies 
for circumstances where \cbr fails to produce a suitable case.
These fallback solutions are also modeled by domain experts and thus explicitly 
tailored to the underlying \cpps. 

\subsection{\cbr Modeling Languages} 

\begin{figure}[htb] \centering
	\includegraphics[width=\columnwidth]{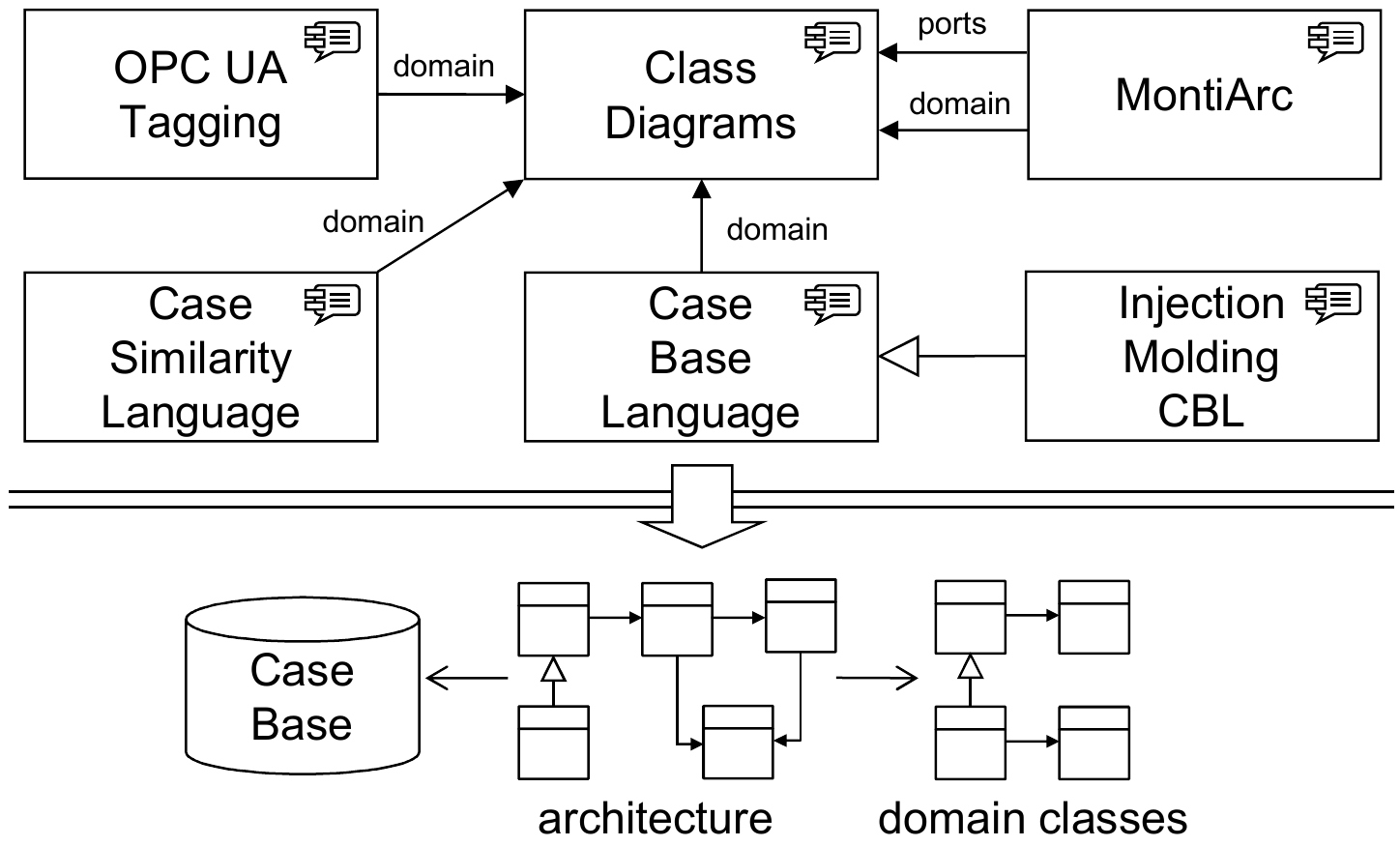}
	\caption{Modeling languages, relations, and artifacts specifying the 
		domain, cases, and fallback
		option for an application of the \cbr framework.}
	\label{fig:cbr_languages}
\end{figure}

Modeling languages facilitate the specification of the \cbr framework and
assist domain experts in making their expertise machine-processable.
Their models tailor the steps of the \cbr cycle and the case base to a
specific application domain.
\autoref{fig:cbr_languages} gives an overview of the integrated modeling
languages employed in our approach: 
\begin{inparaenum}[(1)]
	\item Class diagram models describe the elements and relations of the domain
	and specify data structures available to the framework.
	\item Case base models describe acquainted cases of the physical system. 
	The framework interprets and synthesizes case models at runtime.
	\item Case similarity models specify how to compute the similarity between 
	cases 
	based on their attributes.
	\item Models of the MontiArc architecture description language define the 
	components and architecture of the \dt implementing the \cbr loop. These 
	are predefined and provided with the \dt framework.
	\item OPC UA tagging models~\cite{BDH+20} define how the \dt architecture 
	connects to the API of the \cpps.
\end{inparaenum}

The case base language foresees extension with domain-specific expressions
and actions using the language extension mechanisms of
MontiCore~\cite{HR17}.
The code generated from the case similarity models supports integrating
handcrafted code to define more complex similarity analyses.

\noindent\textit{Class Diagrams}
\vspace{.2em}

\begin{figure}[htb] \centering
	\includegraphics[width=\columnwidth]{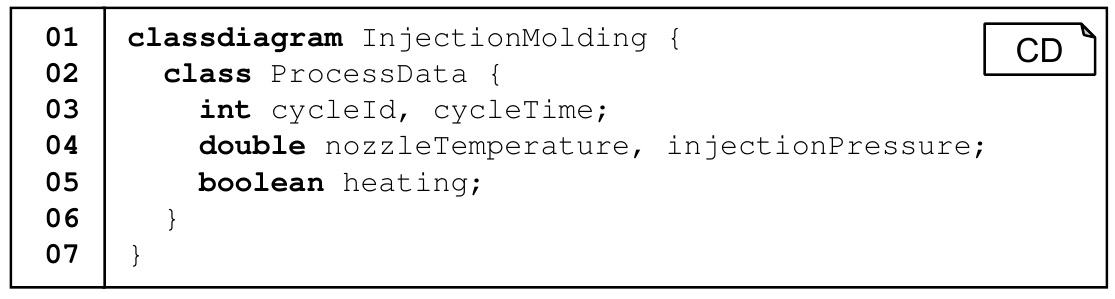}
	\caption{Class diagram for an excerpt from the domain of injection
		molding, containing a small set of parameters of the injection process.}
	\label{fig:cd_example}
\end{figure}

\noindent\cds describe the elements and relations of the application domain.
The case-based reasoner utilizes the corresponding data structures to build
and compare cases.
\autoref{fig:cd_example} presents a textual UML/P \singular{CD} that
illustrates an excerpt of the domain of injection molding. Class
\cw{Process\-Data} symbolizes a data record in the molding process. Besides
metadata like the \cw{cycleId} and \cw{cycleTime} (l.~3), it provides the
values of the nozzle temperature and the injection pressure (l.~4).


\noindent\textit{Case Base Language}
\vspace{.2em}

\noindent The case base language supports domain experts in defining cases.
To this end, they distinguish between \emph{known} and \emph{unknown}
cases.
Known cases describe undesired situations for which the domain expert
knows a solution with its expected consequences.
If a similar situation occurs, one or multiple solution steps can be
repeated to solve the problem.
Unknown cases describe undesired situations for which the expert does know
that the situation might occur and that the system configuration has to
adapt, but does not know how to adapt it.
Although domain experts might not be able to provide precise instructions, they
can provide helpful context knowledge to sort out the problem through a
fallback system.
By combining both types of cases, the domain expert can describe the whole
space of situations that can occur and should be handled by a \cbr system.

\begin{figure}[htb] \centering
	\includegraphics[width=\columnwidth]{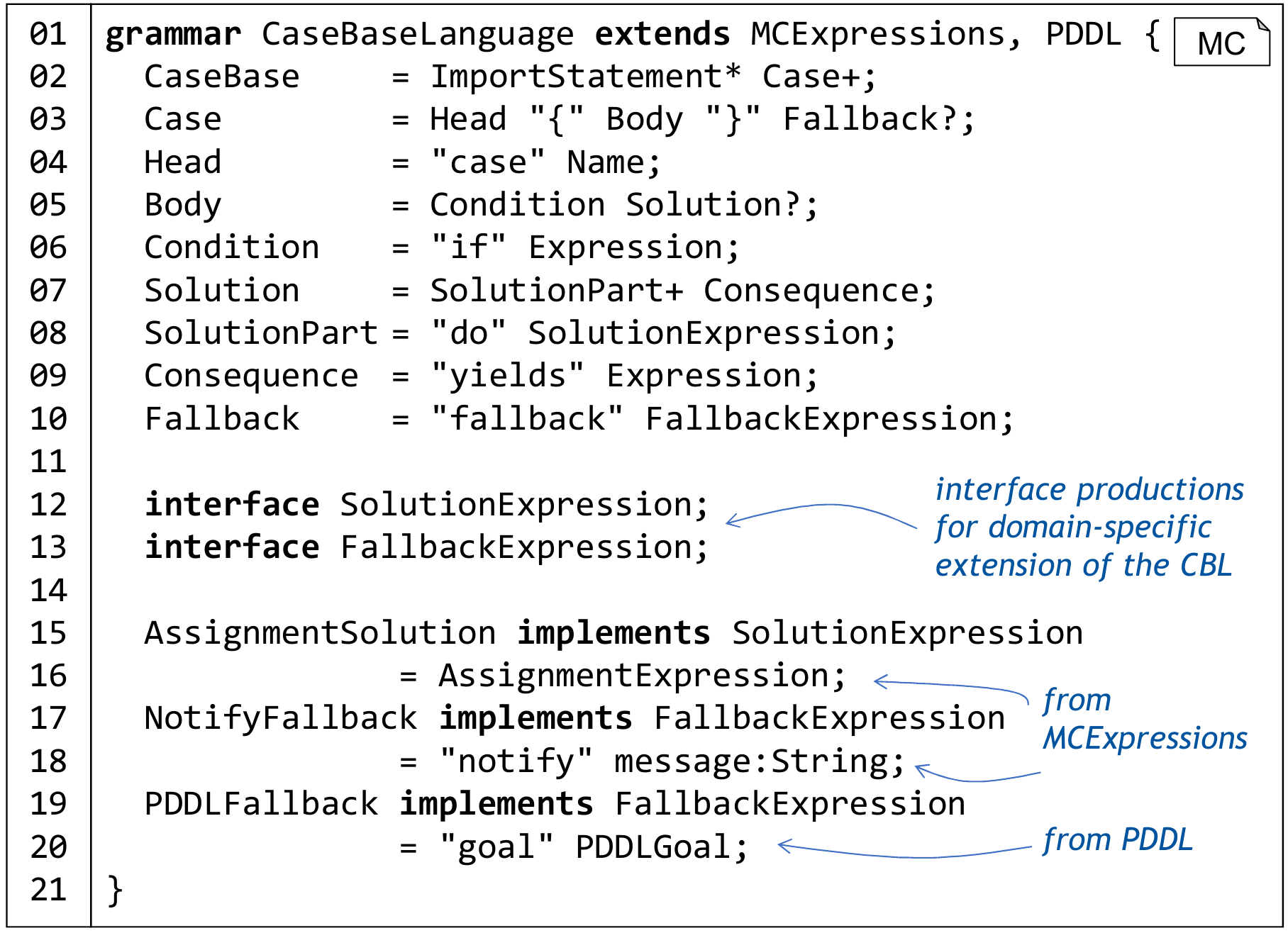}
	\caption{Excerpt of the \mc grammar of the case base language.}
	\label{fig:TB_Case_Base_Grammar}
\end{figure}

The \cbl is defined as textual MontiCore modeling language (\cf
\autoref{fig:TB_Case_Base_Grammar}).
Every case definition references the domain CD by importing models (l.~2)
using the non-terminal \cw{ImportStatement} provided by inheriting from
MontiCore's \cw{MCExpressions} language for binary expressions, statements,
and types. Each case base can contain multiple cases (l.~2) and each case
consists of a head, a body, and an optional fallback (l.~3).
Its head denotes the state of the case and specifies its name (l.~4).
The body comprises a condition and an optional solution (l.~5).
If the body features a \cw{Solution} part, the case is known.
Otherwise, it is treated as an unknown case.
The condition essentially is a Boolean expression (l.~6) over any types and
properties available through the domain model.
Well-formedness rules of the \cbl ensure that the expressions are valid
Boolean expressions (\ie referenced types exist and can be compared as
specified by the expression).
The solution is a non-empty sequence of solution parts with a consequence
(l.~7).
Each solution part (l.~8) refers to the interface non-terminal
\cw{SolutionExpression} (l.~12) that is an extension point of this grammar
and can be implemented in domain-specific sub-grammars.
Per default, arbitrary assignments are supported as solutions (ll.~15-16)
and corresponding well-formedness checks are provided.
Further, domain experts can also specify java code that performs a solution and call this code in the solution part.
Java calls are realized by the imported \cw{MCExpressions}. 
The \cbl also supports PDDL specifications (provided by importing \cw{PDDL}) for planning if the solution for a case is 
unknown. 
The consequence describes a postcondition that should hold after the case
has been applied. The \cbr system relies on the specified \cw{yields} consequence to check whether a case was successfully applied. 
If the specified postcondition is not fulfilled, the \dt applies this case less likely in the future.
Similarly, the fallback (l.~10) is an extension point for fallback actions
to be used if the case fails.
Per default, notifying users (ll.~17-18) and falling back to PDDL planning
(ll.~19-20) are supported.
Both solutions and fallbacks are meant for extension through
domain-specific or application-specific sub-grammars in which, \eg temporal
expressions, fallback automata, or other means can be integrated using
MontiCore's language extension mechanisms~\cite{HR17}.

\begin{figure}[htb] \centering
	\includegraphics[width=\columnwidth]{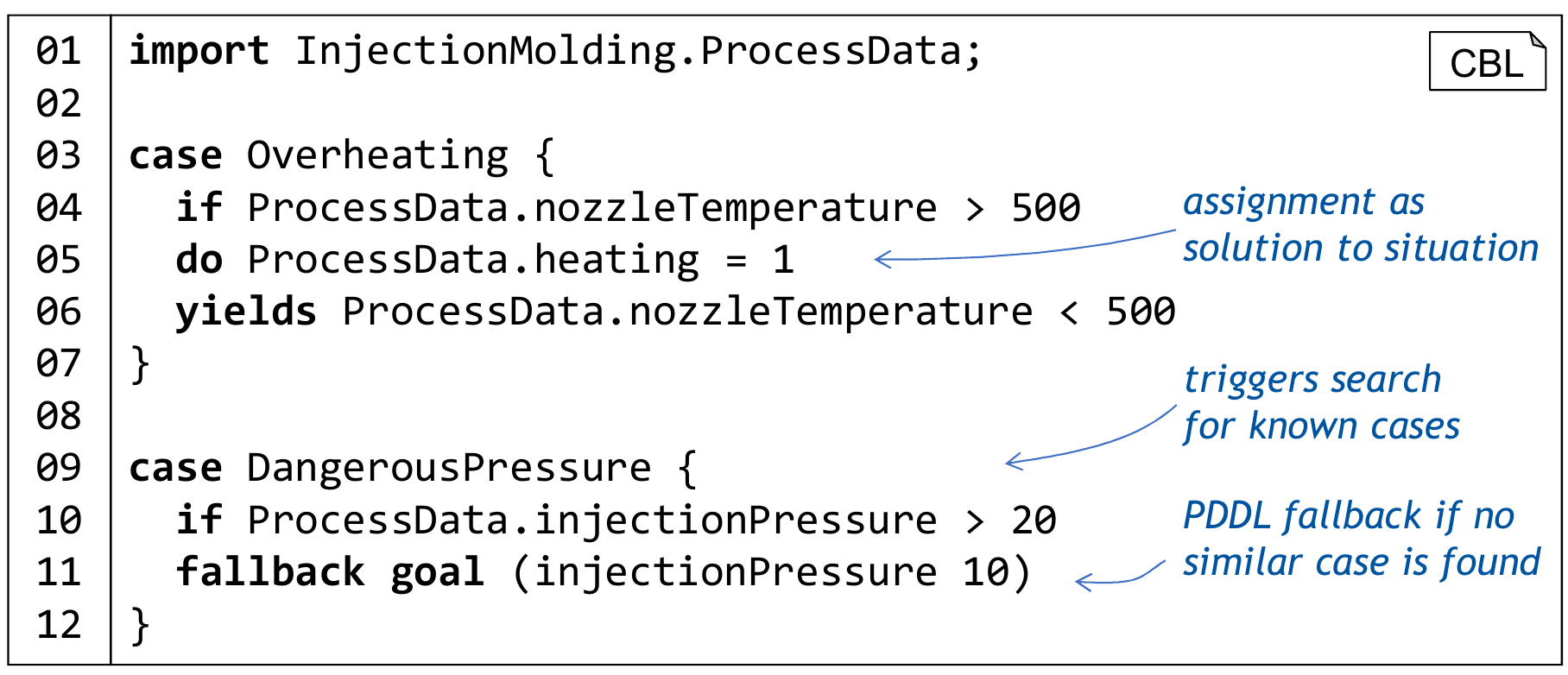}
	\caption{Excerpt of a case base for injection molding regarding dangerous
		temperatures and pressures.}
	\label{fig:case_base_lang_model}
\end{figure}

\autoref{fig:case_base_lang_model} illustrates a model of the \cbl that
refers to the injection molding domain model \cw{Process\-Data} (l.~1) and
shows two cases that may occur in the injection molding machine.
The first case represents a known case (ll.~3-7) that handles a problematic
temperature of the injection nozzle characterized by being higher than $500$
degrees Celsius (l.~4).
The attributes in this expression reference the domain model of
\autoref{fig:cd_example}.
As a solution, the case contains an assignment expression that specifies
setting the \cw{heating} to level \cw{1}.
The second case is unknown (ll.~9-12) and addresses dangerous pressure in
the injection process~(l.~10).
When the condition holds, a retrieval of similar known cases is triggered.
In case the search yields no cases, the \dt uses the PDDL fallback expression
to start finding a plan over the \cpps actions and properties that, when
executed, will reduce the pressure.

\noindent\textit{\glsdesc{CSL}}
\vspace{.2em}

\begin{figure}[htb] \centering
	\includegraphics[width=\columnwidth]{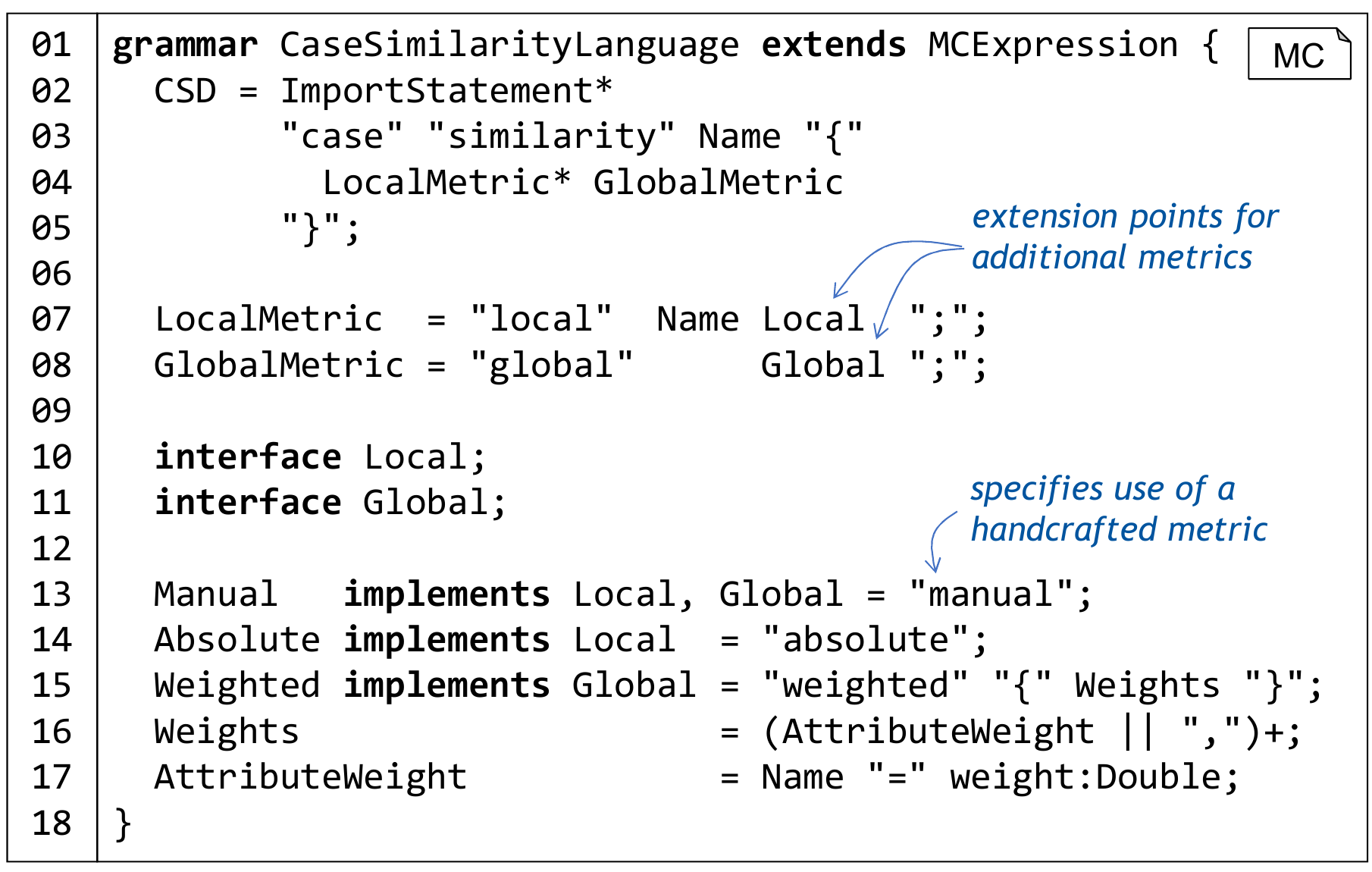}
	\caption{Excerpt of the \mc grammar of the \csl.}
	\label{fig:case_similarity_language}
\end{figure}

\noindent The second essential part of a case-based reasoning system is its
ability to assess similarity between a situation in the \cpps and a case.
Similarity as a metric is expressed as a positive rational number with
\cw{0} being considered equal.
The \csl supports describing weighted global and local similarity based on a
the types of domain models and promotes integration of further, handcrafted,
similarity analyses using the top mechanism~\cite{HR17} with its generated
code artifacts.
We developed a \dsl for specifying 
similarities between cases (\autoref{fig:case_similarity_language}). 
Every case similarity definition is based on domain models and consists of
global and local similarity metrics (ll.~2-5).
After an import list establishing relation to the domain of
discourse~(l.~2), a name case similarity specification follows (ll.~3-5),
which contains (l.~4) multiple local similarity metrics (l.~7) that relate
to individual attributes of the domain models and a single global similarity
metric (l.~8) describing how these individual similarities are weighted.
Both kinds of metrics reference to interface non-terminals (ll.~10-11) that
facilitate introducing new metrics into the \csl.
The \csl features two kinds of metrics for local and global similarities
(ll.~13-17) out of which the \cw{manual} metric specifies that a handcrafted
similarity analysis should be used.
This demands implementing a specific Java interface of the
\csl's runtime system, which is then invoked if the \cw{manual} is used.
Well-formedness rules ensure that the referenced domain types exist and are
correctly used.

\begin{figure}[htb] \centering
	\includegraphics[width=\columnwidth]{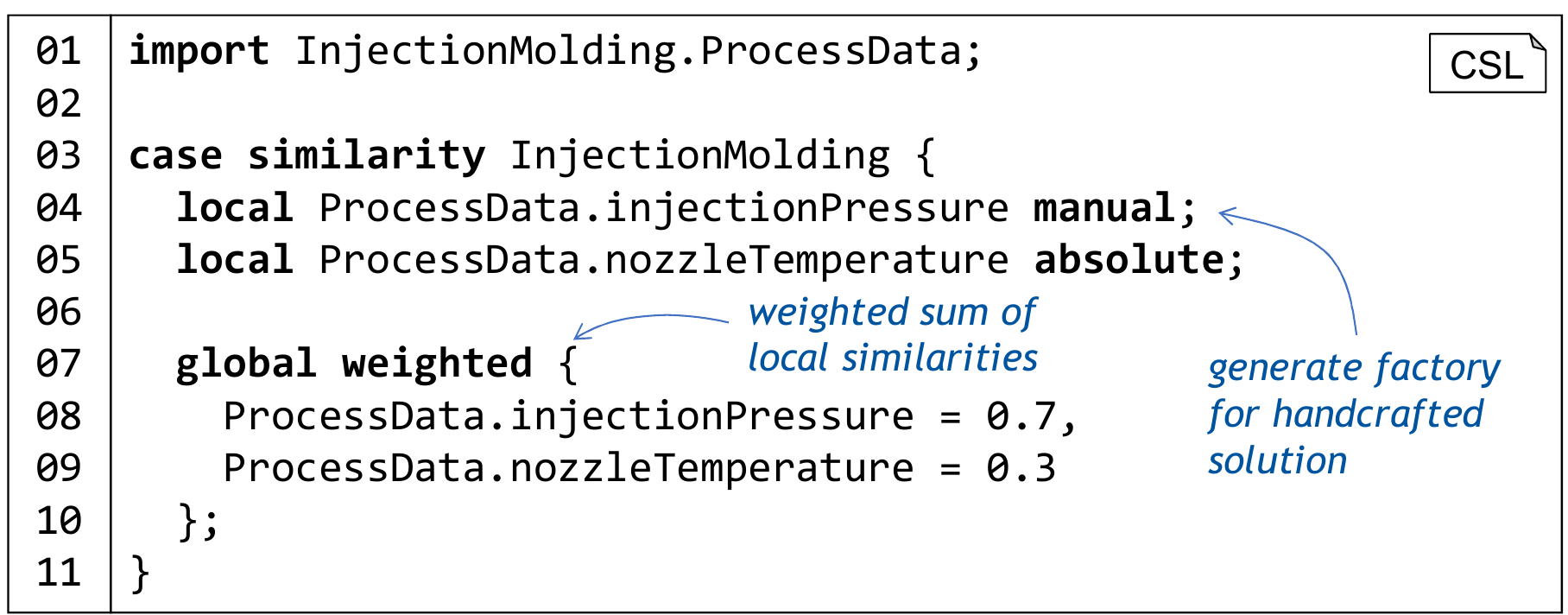}
	\caption{ Exemplary case similarity model. }
	\label{fig:case_sim_lang_model}
\end{figure}

\autoref{fig:case_sim_lang_model} displays a model of the \csl.
It refers to the injection molding domain model \cw{ProcessData} (l.~1) and
specifies two local similarities for this domain (ll.~4-5).
The model specifies the absolute distance local similarity metric for the
\cw{nozzleTemperature} and a handcrafted metric for the \cw{pressure}
attribute.
Global similarity then is defined through the weighted combination of
\cw{nozzleTemperature} and \cw{pressure}~(ll.~7-10).

Similar to the \cbl, the \csl also does not aim to be a catch-all language
but supports leveraging MontiCores language extension mechanisms to define
highly specific similarity metrics (\eg featuring uncertainty, SI units, or
domain terminology). 


~\\\noindent\textit{\pddl Fallback Modeling}
\vspace{.2em}

\noindent The selection of cases to resolve a situation bases on their
similarity.
Depending on the size and scope of the case base, no case might be
available.
As this is not unusual, the \dt has to be able to handle such situations.
Therefore, we provide a fallback possibility to full AI planning.

Due to the dependence on the domain, we do not prescribe a specific modeling
language for fallback activities but provide extension points in both \cbl
and \dt.
The default implementation of these extension points are notification of human
operators and invoking a PDDL planning.
Depending on the case defined by the domain expert, parameters for a
\cw{Fallback} component or \pddl goals can be defined.
For the former, a \dt component taking care of the fallback activities has
to be provided, for the latter, the corresponding goal must be defined in
the \dt's \pddl knowledge base.



\subsection{Integrating Domain-Specific Models into CBR}
\label{sec:cbr_workflow}

In our approach, the \csl models are interpreted at \dt runtime, whereas the
\csl models are used for code generation at design-time to enable the
integration of handcrafted, more complex, similarity analyses (\cf
\autoref{fig:framework_architecture}).
This section explains their overall integration.


~\\
\noindent\textit{At Design-Time}
\vspace{.2em}

\noindent The similarity measures do not change once the \dt is running.
Hence, at design-time, \csl models are translated into Java artifacts that
are invoked when their computations are necessary.
As the models feature planned extension with handcrafted Java computations
(indicated by the \cw{manual} keyword), we exploit the code generation to
support injection of these handcrafted similarity analysis artifacts using
the TOP mechanism~\cite{HR17}, a variant of the generation gap pattern, and
generate factories to inject implementations of known similarity analysis
interfaces into the similarity computations.
Leveraging Java for more complex analysis liberates domain experts
specifying similarities from the complexities of an overly generic (possibly
Turing-complete) modeling language and supports engineers in using
established tools, frameworks, and libraries to develop the analyses.
Where more complex analyses are required within in the \cbl, MontiCore's
language extension enables creating sub-languages whose grammars implement
and extend the \cbl's extension points. 


Each model for local similarity is translated to a class with a name of the
form \cw{<Name>LocalSimilarity}.
\cw{<Name>} is replaced with the name of the domain model attribute.
The class provides a single public method to calculate the local similarity.
As a parameter, the respective attribute and the condition expression of the
case to compare to are given.
The domain model determines the type of the attribute.
Therefore, type-safe artifacts can be produced, and their stable interfaces
hide whether these are generated or handcrafted from the framework.

The model of global similarity results in a class with a name of the form
\cw{<Name>Similarity}, where \cw{<Name>} is the name of the overall similarity
model (\autoref{fig:case_sim_lang_model}, l.~1).
The class provides a single public method to calculate the global
similarity.
All domain attributes and all condition expressions are passed as
parameters.
Based on the defined similarity type, an implementation is generated.
This holds for the global weighted similarity.
The \cbr modeling framework collects the artifacts for the local
similarities based on the models. Based on the weights, their calculated
similarity is summarized.

Ultimately, this enables domain experts to develop their own \dts and enrich 
these with domain knowledge without requiring any programming skills. 


~\\
\noindent\textit{At Runtime}
\vspace{.2em}

\noindent The case base comprises cases, which are interpreted at runtime.
To this end, they are parsed, and their abstract syntax representation is
stored in memory.
During runtime, the \dt monitors the \cpps and checks for undesired
situations using the case models.
If such a case is found, \dt action is required, and it tries to retrieve the
most similar case.

During the \emph{retrieve} phase of the CBR cycle, the \dt
receives the undesired (current) situation and the list of known cases as
input.
To determine similar cases, the similarity is calculated for every case.
The metric value is determined based on the conditions and the situation.
This step relies on the similarity computation artifacts generated based on
the \csl models and the related handcrafted artifacts.
Next, the results are filtered by a predefined constant threshold.
Similarities between cases range between 0 and 1 in our implementation, and we 
consider a value smaller than 0.2 as similar enough to try to apply the 
solution of a case. 



In the reuse phase, the \dt then determines the actual solution to execute.
For this, the previously selected set of similar cases is taken.
By default, the \dt tries to employ the most similar case.
How new cases should be constructed and under which assumptions their
solutions can be synthesized again is highly domain-specific and depends on
the context our framework is employed in and the connected \cpps.
For instance, synthesizing multiple new cases to experiment with finding
\cpps behavior optima might be a valid approach in an initial deployment
setting but not during normal operations.
Hence, our framework supports extension with more sophisticated reuse
mechanisms, such as constructing new cases by deviating case conditions and
solutions systematically or interpolating between multiple similar cases.

After executing a solution, the \dt uses the retain phase to learn from the
result, which requires the situation before and after applying the solution.
Based on the situation after executing the solution, the expected outcome is
compared to the resulting of applying the solution.
If the resulting outcome matches the expectation, the existing case is
either reinforced as being useful or the new case is added to the case base.
For the latter, the situation's properties are therefore converted into
equality equations.
Next, the similarity of the new case to those in the case base is assessed.
If the smallest similarity is above a domain-specific learning threshold,
the \dt considers the case as new and adds it to the case base models.
Independent of whether a new case was learned, the situation triggering the
\cbr, the selected cases, solutions, and outcomes are logged for the
operators to support explaining system behavior.

\section{Model-Based Framework for Case-Based Reasoning}
\label{sec:Framework}

\begin{figure}[htb] \centering
	\includegraphics[width=\columnwidth]{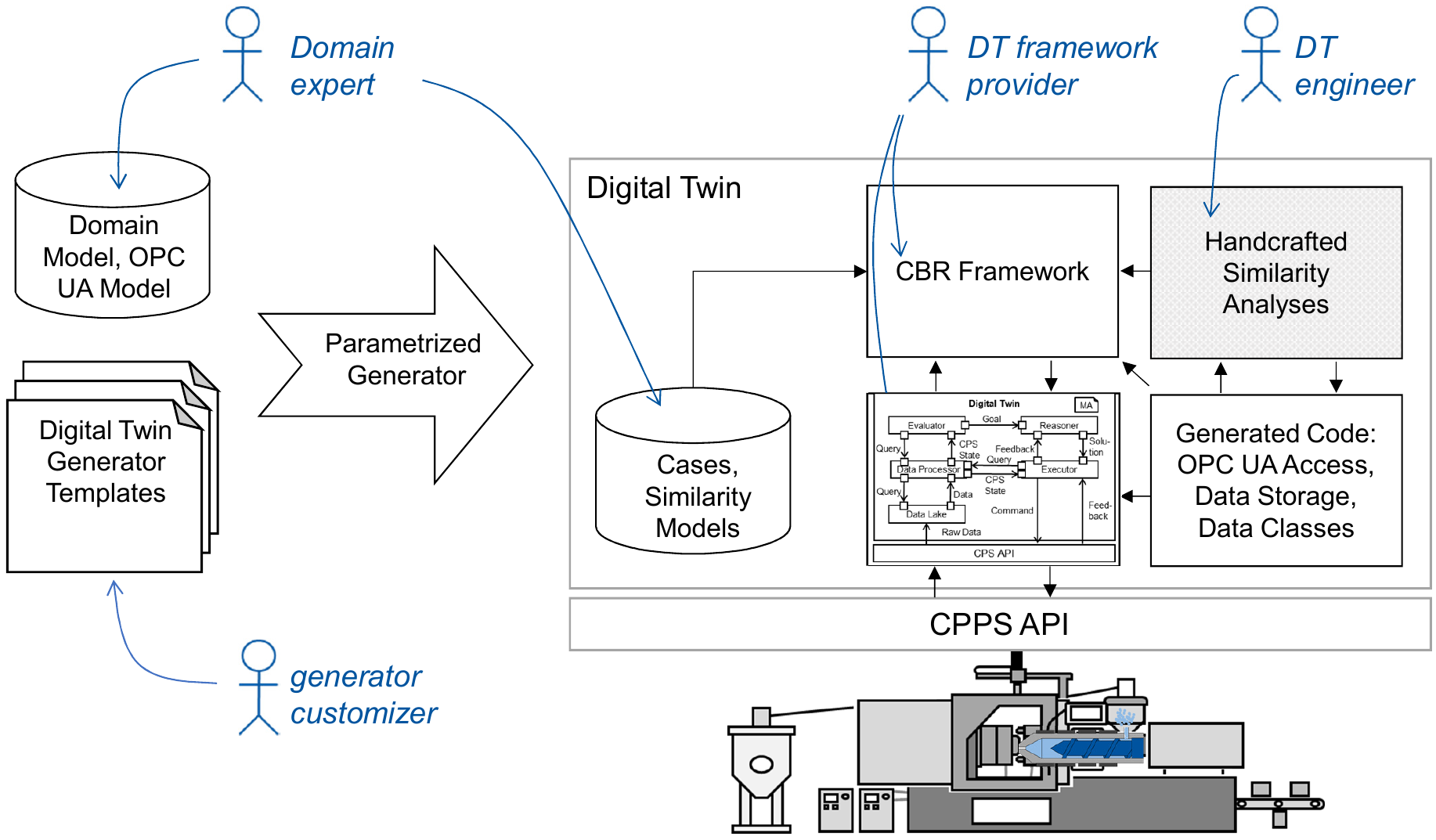}
	\caption{Framework for generating a \dt based on domain knowledge provided in models.}
	\label{fig:framework_architecture}
\end{figure}

By providing \dsls and adequate code generators, we enable domain experts to adapt 
the \dt framework we built to individual \cpps and specific requirements. 
\autoref{fig:framework_architecture} presents the realized framework for
generating \dts.
It contains the general \dt services for storing data, sending OPC UA
commands and evaluating data to identify the current system state.
Furthermore, it offers the general functionality to perform \cbr.
The components for assessing the current \cpps state and storing data 
are predefined in the framework and tailored to the application scenario by
generating, \eg the actual data base structure and OPC UA 
commands according to the domain and OPC UA models. 
Since all parts of the \dt can be generated, explicitly no software developers 
are required to create a \dt.
The generator creates a \dt that is self-adaptive based on the domain knowledge
that is provided as cases. 
If the \dt detects an unintended behavior, it adapts the configuration of the 
physical twin accordingly. If this does not improve the \cpps's behavior
and results in the state described in the case's consequence part, the \dt learns that the case is not successful and tries to apply an alternative.  

The \dt is tailored to a specific \cpps through models, describing this
\cpps.
A domain expert specifies the \cpps in a domain model and adds information
for data retrieval via an OPC UA tag model.
These models serve as input for the generator that creates Java code for OPC
UA access, data objects, and storage.
To enrich the generated \dt with domain knowledge for self-adaptation, the
domain expert also creates case models that characterize critical situations
at runtime and how the \dt should handle these situations. Besides, the
domain expert specifies similarities to determine whether the \cpps
situation at hand resembles one of these cases.
The case and similarity models are interpreted while the \dt is running.
Thus after generation, the domain expert can add further cases to the case
base if necessary.
The \dt calculates the similarity of the situation in the \cpps and a case in the case base by mapping the actual machine values with parameters in the case. 
Since the case and the data access are consistent with the domain model (\cf \autoref{fig:cd_example}), the \dt can map sensor values with parameters in the case. 
\eg the current value of the \cw{nozzleTemperature} (l.4) is sensed by a sensor in the machine and mapped to the parameter in the case (\cf \autoref{fig:case_base_lang_model} (l.4)) when the \dt calculates the similarity. 

\begin{figure}[htb] \centering
	\includegraphics[width=\columnwidth]{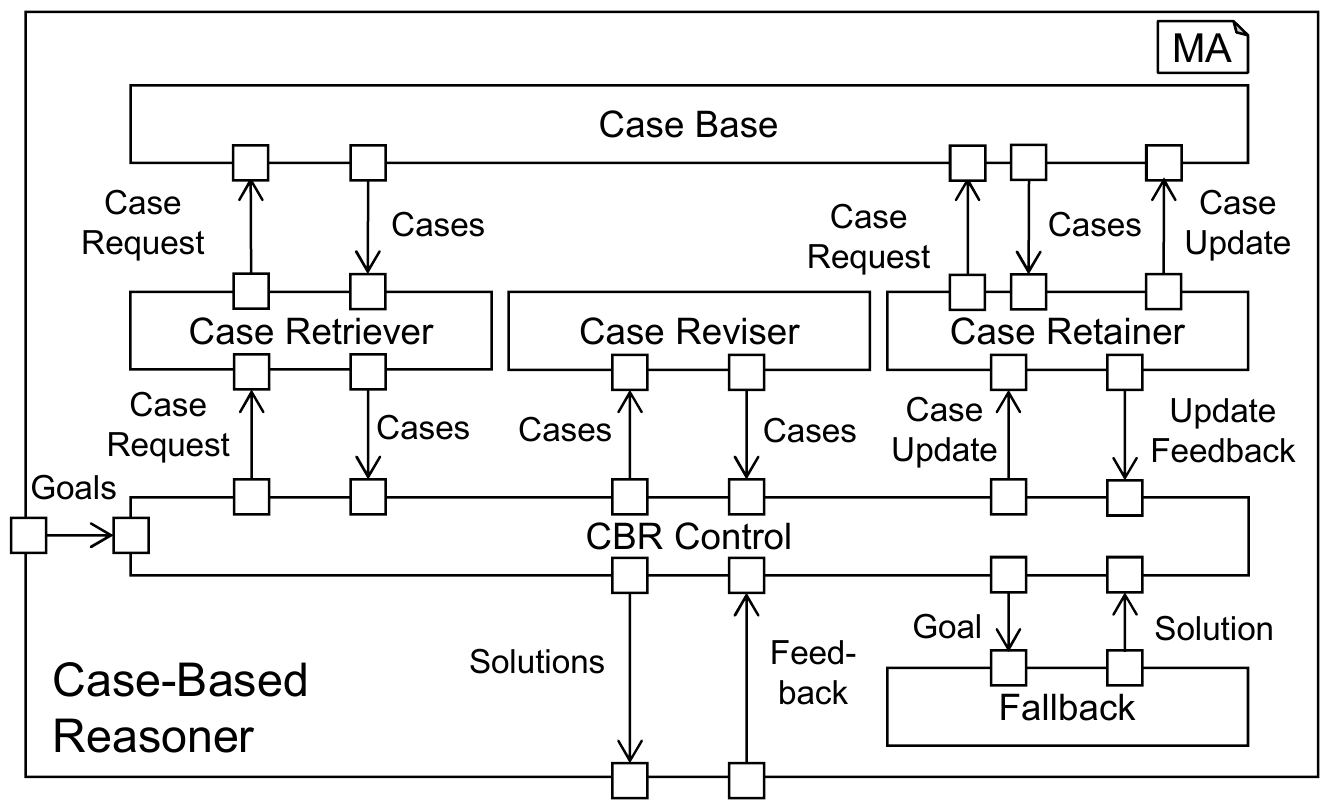}
	\caption{Internal composition of the \cw{Case-Based Reasoner}. It connects to 
		the \cw{Case Base} and comprises components for the steps of the \cbr cycle as well as a fallback. }
	\label{fig:framework_reasoner}
\end{figure}

Internally, the \cw{Case-Based Reasoner} comprises six sub-components that 
are
responsible for the individual \cbr activities (\cf \autoref{fig:framework_reasoner}).
A control component manages the process and interacts with the respective
\cbr components. The \cw{Case Retriever} obtains those cases from
the \cw{Case Base} that are similar to the current problem situation.
The \cw{Case Reviser} tailors the contained solutions to the problem at
hand. Additionally, it reacts to feedback received from the Executor and
further adapts the solution if necessary. 
When the \cw{Case Base} does 
not contain known cases, the \cw{Case Reviser} employs the
\cw{Fallback} which is usually notifying the machine operator or stopping the machine. 
Finally,
the \cw{Case Retainer} stores the experience, including the encountered
the problem, applied solution, and its success, in the \cw{Case Base}.

We implement the \cbr framework (\ie the
\cw{Case-Based Reasoner} and its sub-components) as an extension for the base \dt architecture (\cf \autoref{fig:MD_DT_Architecture}). For that
purpose, we provide a general implementation for the \cbr components and define
the domain-specific details via \cbr models.
Models of the Case Base Language describe known cases in the domain at hand
and, thus, determine the Case Base contents and guide a system's management.
The \cw{Evaluator} monitors the system by checking the occurrence of unknown
cases. The \cw{Case-Based Reasoner} utilizes the known cases and the
Fallback to find a solution for the detected situations. To that end, it
relies on the Case Similarity Language models to determine the similarity
between a case and the given situation to find an applicable case or
adequately store new experiences. Fallback models provide an alternative
method of solution-finding when \cbr does not yield a suitable solution.
The generated \dt relies on this framework when performing self-adaption through \cbr
but is enriched with domain-specific models that experts can provide. 
\section{Application Example}
\label{sec:Example}

We created a \dt with \cbr for an injection mol\-ding machine as a demonstrator.
The \cbr framework and the \dt architecture were specified by us while domain experts from injection molding created models of the \cbl and \csl. 
We tested the generated \dt on real data from a filling experiment series in injection molding. 
After mounting a new mold part for series production, the exact parameter
settings are unknown, and the operator usually runs a so-called filling study to 
slowly approach an ideal
configuration.
Step by step, the amount of injected plastic is increased until the mold is
filled.
Then, fine-tuning finds a configuration that also ensures a smooth surface 
of the part and reduces leakage.
We aim to speed up the process of finding the correct parameters while 
focusing on one specific machine and one mold.

Our \dt is tailored to a ALLROUNDER 520 A 1500 by ARBURG.
The adaption efforts can roughly be structured as follows:
1) provide data binding to the machine,
2) identify domain model,
3) devise a case base, and
4) establish a notion of similarity.
For data access, the manufacturer provides an OPA UA interface through which the \dt can access runtime data. 

\begin{figure}[htb] \centering
	\includegraphics[width=\columnwidth]{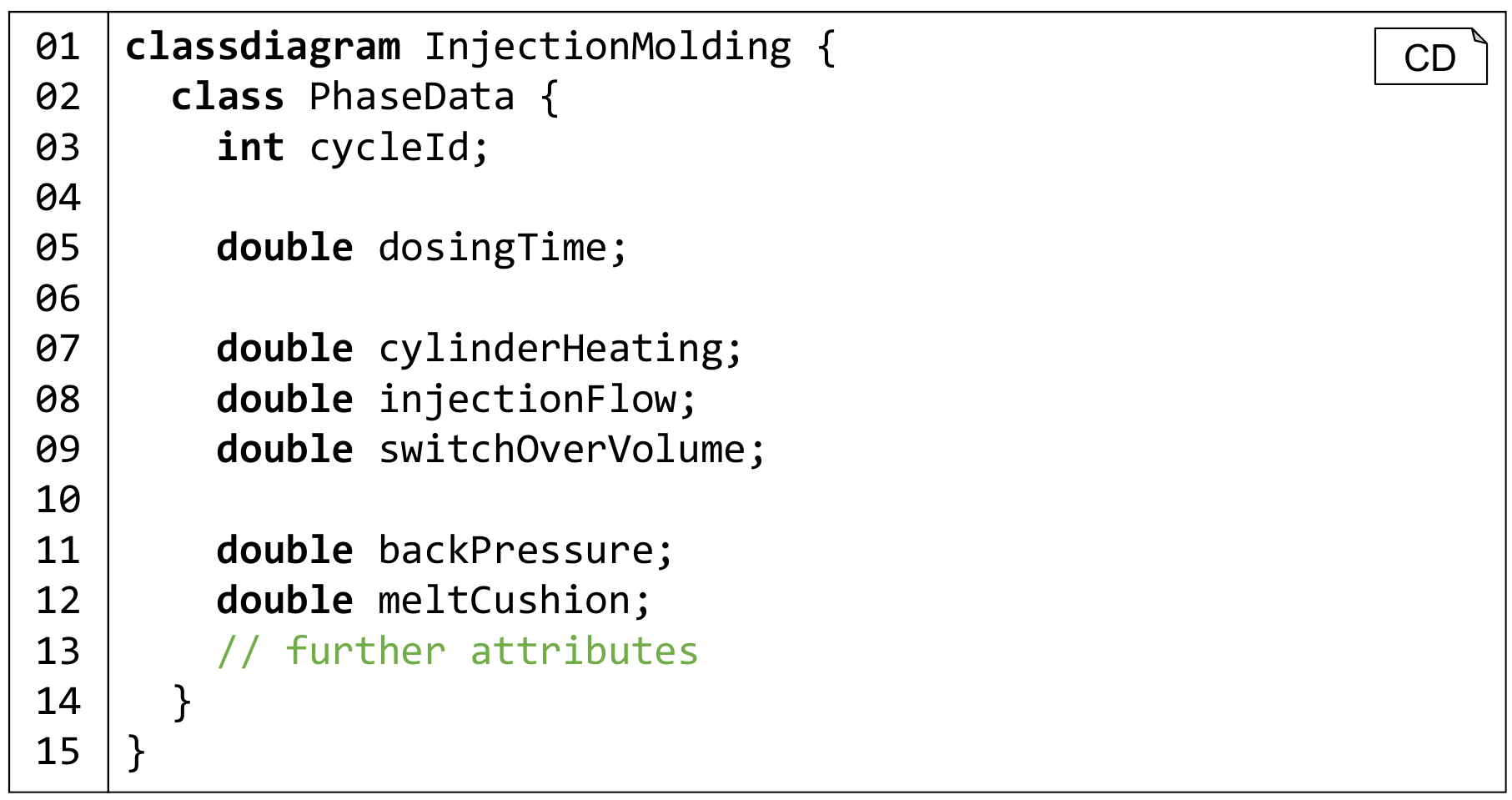}
	\caption{
		Injection molding domain model.
		\cw{PhaseData} comprises parameters of the production of a single part.
		Only the most critical parameters are depicted.
	}
	\label{fig:study_domain_model}
\end{figure}

In cooperation with domain experts from injection molding, we identified 
representative parameters (\cf \autoref{fig:study_domain_model}) for capturing the machine's state. 
\cw{PhaseData} comprises all parameters of an injection process.
\cw{DosingTime} determines for how long plastic is loaded into the
plasticizing unit.
The attributes \cw{cylinderHeating}, \cw{injectionFlow}, and
\cw{switchOverVolume} (ll.~7-9) describe the injection parameters.
\cw{cylinderHeating} sets the temperature inside the plasticizing unit,  \cw{injectionFlow} and \cw{switchOverVolume} 
influence how fast and long plastic is injected.
The \cw{meltCushion} (l. 12) is the surplus of material left in the
plasticizing unit after the injection.
\begin{figure}[htb] \centering
	\includegraphics[width=\columnwidth]{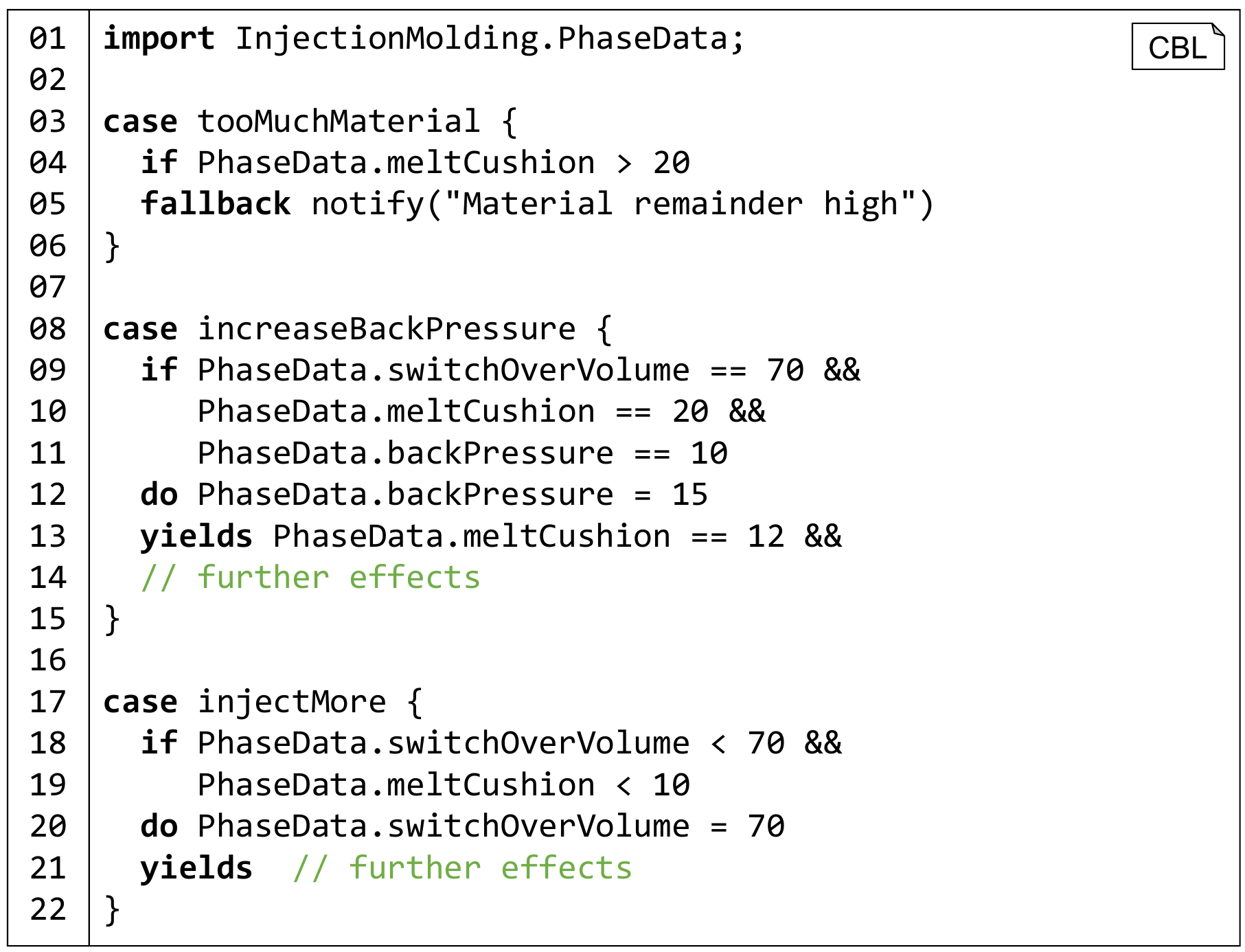}
	\caption{
		Example cases from the case study in injection molding.
		The first case defines the problematic parameter space.
		The others are solutions to handle more specific situations.
		Repeating imports are omitted.
	}
	\label{fig:study_cases}
\end{figure}

\autoref{fig:study_cases} exemplarily shows cases identified for the case study.
The first case is unknown (ll.~3-6).
It describes the problematic parameter state of having too much residual material.
The other two cases (ll.~8-21) feature possible solutions.
The first handles a situation where too little material is injected.
The specified \cw{backPressure} is too low, leading to missing material in the mold.
The second covers the situation where more material can be injected.

\begin{figure}[htb] \centering
	\includegraphics[width=\columnwidth]{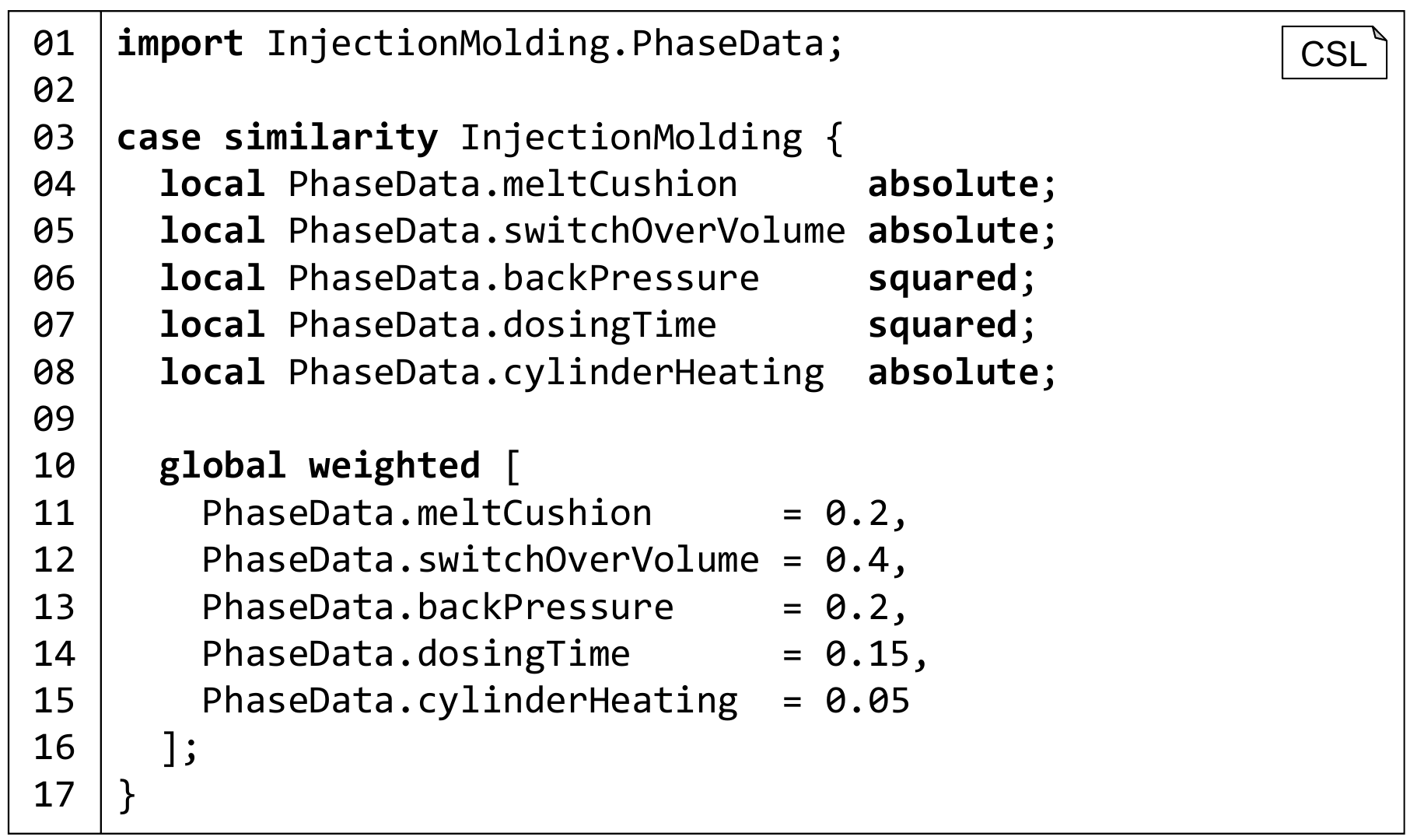}
	\caption{
		Model for similarity calculation in injection molding.
	}
	\label{fig:study_similarity}
\end{figure}

For similarity, we employ a weighted similarity calculation, as specified in
\autoref{fig:study_similarity}.
The local similarities define the critical parameters with influence on the
metric (ll.~4-8).
The values of \cw{backPressure} and \cw{dosingTime} are more
sensitive to changes.
Therefore, we use a \cw{squared} local similarity for them.
It is offered by the \csl where the difference is squared.
The global similarity characterizes the weights (ll.~10-16).
The similarity of \cw{switchOverVolume} has the most influence with a
weight of \cw{0.4}.
A minor influence has the similarity of \cw{cylinderHeating} with a
weight of \cw{0.05}.

Using models and the customized \dt components, one receives a fully functional \dt generated based on domain models, case 
models, and similarity models to provide a \cbr for an injection molding machine. 

For evaluating the generated \dt, we captured the cycle-times for \cbr cycle execution in the \dt while it was running on a 
local computer with Intel(R) Core(TM) i7-7600U CPU. 
The \dt operated on real historical data of the injection molding machine but, due to safety issues, could not change settings 
on the machine.

Initially, the \dt started with a case base of $20$ cases that we identified in cooperation with the injection 
molding experts.  
The measured results are displayed in \autoref{tab:runtimes}.
\begin{table}[!h]
	\centering
	\caption{Cycle times of the \dt's \cbr cycle.}
	\begin{tabular}{lccc}
		&  \textbf{Minimum} & \textbf{Maximum} & \textbf{Average}  \\ 
		&  \textit{ms} & \textit{ms} & \textit{ms}  \\ \midrule
		\textbf{First Cycle}	& $16,9181$ & $110,2881$ & $42,2136$  \\
		\textbf{No Case}	& $2,6234$ & $50,6226$ & $16,16197$  \\
		\textbf{Case Detected}	& $1,7137$ & $75,6592$ & $13,01902$ \\ \toprule
	\end{tabular}
	
	\label{tab:runtimes}
\end{table}
The \dt's \cbr cycle was triggered every time that the machinecycle counter in the machine changed. This parameter simply counts the number of performed production cycles on the ARBURG.
During the first cycle, the \dt loads the initial cases from the file system. Consequently, this cycle's duration had a 
longer 
execution time ( $42,2136$\textit{ms} in average) than other cycles.
The \dt monitors the injection molding machine and compares the current state to identified cases. 
When no case matched the current machine's state, this comparison took $16,16197$\textit{ms} on average. 
If a case was present, the \dt detected it within $13,01902$\textit{ms} on average and tried to adapt its behavior based on the solution stated in the case model.  
If the machine data confirmed the case's success, the \dt marked the applied case as successful or unsuccessful, 
respectively. 
We expect an increase in cycle time due to communication latency if the \dt connects to the \cpps and autonomously 
changes process parameters.
Given that injection molding is a cycle-based process where the process settings can only be updated for the next cycle, and 
that production cycles take between 50 seconds and 2 minutes, the computing times of the \dt are sufficient to adapt the 
process settings in time. 
\section{Discussion}
\label{sec:Discussion}

We applied the presented framework and modeling languages to create a \dt of
an injection molding machine. The realized \dt establishes a connection to
the injection molding machine and reads its sensor values.
Based on these, the \dt autonomously detects unintended system behavior and
produces solutions based on similar cases provided by the case base.

While cases consisting of conditions and effects are very intuitive, the
modeling languages of our framework rely on some experience with data types
and structures (int, float, Boolean, objects), an understanding of model
relations (imports), and might even relate to PDDL knowledge bases.
The first challenge can be mitigated by providing even more
domain-specific extensions of the CBL that rely only on data types and data
structures well known by the domain experts and by intelligently translating
these to the data structures communicated via OPC UA to the \cpps.
The notion of model imports could be omitted by fixing a \cbr \dt to a
single domain model class diagram and adjusting the CBL again.
Similarly, PDDL fallbacks can be prohibited in domain-specific sublanguages
of the CBL.
Hence, the languages employed within our framework can be tailored precisely
to the complexity suitable for the domain experts operating the systems.
These, of course, limit the usefulness of the overall framework. 
Nonetheless, due to MontiCore's language extension mechanisms, making the
language as comprehensibly as necessary is possible (\textbf{R1}). 
In general, \cbl and \csl were regarded as easy to understand and use. 
However, injection molding experts had difficulty in explicitly expressing similarities 
between cases because they often also work by gut feeling and could not pinpoint the 
exact point that triggers their adaptation of the machine configuration.

The generated \dt works autonomously (\textbf{R2}) and evaluates the current 
\cpps state every time that a new production cycle starts. 
If the \cpps state matches the condition of a case, the \dt adapts the \cpps 
configuration based on the solution specified in the case. If this adaptation 
does not lead to the expected behavior of the \cpps the \dt learns 
to prioritize this case lower in the future.

The presented \dt can connect to any \cpps that provides a communication
interface; thus, it receives the data for evaluating if any unintended
situations occurred.
Active writing of parameters to the machine while it is running remains
critical and, due to liability issues, might be prohibited in other domains.
Nonetheless, the \dt provides solutions for detected cases and attempts to
implement these autonomously. If the connected \cpps prohibits manipulation
of settings without human interaction, the \dt can at least provide a
recommendation for adapting the machine configuration.
Moreover, the presented framework is reusable for other \cpps (requirement
\textbf{R3}) as essential parts of the \dt are modeled independent of the
underlying \cpps. Transferring the \dt to another \cpps requires
implementation of adapters to communicate with the machine,  manipulating the domain model, and specifying an application-specific case base and similarity measurements.
The model-driven development of the \dt based on a generator that derives
the concrete implementation from domain models further speeds up the
development process.
Overall, the various configuration means support tailoring our approach to a
variety of self-adaptive manufacturing scenarios (\textbf{R3}). 

When the \dt that we realized encounters new cases, it first searches
through the case base to find the most similar case and tries to adapt its
solution to the case at hand.
If this adaptation is successful, it creates a new case and adds it to the
case base. Thus, when running over a more extended period of time the \dt
learns more cases and becomes more effective. Thus, the realized \dt
improves over time by persisting experiences that domain experts can review as explanations of self-adaptive behavior (\textbf{R4}).
Interesting challenges arise due to the indeterministic nature of \cpps, the
actions taken in the past may not be relevant for similar cases in the
future. Nonetheless, since the \dt is able to adapt cases in terms of their
success, it at least does not try to apply solutions that verifiably do not
lead to desired situations.

\section{Related Work} 
\label{sec:RelatedWork}

An approach similar to ours utilizes an IIoT Gateway with an OPC UA
interface as a mediator between a \dt and the physical system
\cite{RW_DT_RefArch19}. We suggest exchangeable adapter components for both,
data retrieval and control, supporting a range of different communication
technologies and protocols.
A different idea investigates model-based \dts that support and guide
product development in all phases of the life cycle
\cite{RW_DT_Simulation16}. During design and engineering, \dts comprise
collections of digital artifacts (data and models) to provide simulations of
the expected system behavior. 

A similar concept utilizes \dts to merge
different kinds of system data to model its behavior
\cite{RW_DT_AllPhases18}. Thus, the \dt shows the effect of design changes
on the physical system and supports virtual verification of its behavior.
Further research demonstrates the extent of technologies and application
domains for \dts in manufacturing.
In a framework for smart workshops, \dts control \cpps, providing local
optimizations and communicating to achieve a global optimization
\cite{RW_DT_App_Workshop19}.
Another approach employs edge, fog, and cloud computing to implement \dts
\cite{RW_DT_App_SmartManufacturing18}. The \dts control physical entities
via virtual models and are connected on a network level or through the cloud
to perform optimizations of increasing degree.
However, these proposed \dts are tailored to the given tasks or application
domain while we present a customizable approach that is applicable to a 
wide variety of purposes.

Autonomic system must be able to handle unexpected and novel situations.
Thus, \cbr is well suited for application in autonomic systems and
especially in \dts.
This includes employing CBR for self-configuration in autonomic systems
\cite{RW_CBR_AutonomicSystems1_10}
or
utilizing CBR to detect and repair system failures at runtime (self-healing)
\cite{RW_CBR_SelfHealing2_08}.
These approaches face the cold-start problem, though. As a solution is
derived from existing cases, considerable effort and knowledge about the
domain is required to set up an extensive case base.
A solution to the cold-start problem is a combination of \cbr and goal
reasoning  \cite{RW_CBR_SelfAdaption14, RW_CBR_SelfAdaption15}.
A case-based reasoner tries to solve problems based on the experiences in
the case base. If the cases do not yield a solution, the system applies goal
reasoning as a fallback to create new cases and adds these to the case base.
An alternative is building the case base in an offline learning phase
\cite{RW_CBR_ReinforcementLearning17}.
The approach utilizes reinforcement learning for creating new cases. In the
online learning phase, the system finds appropriate cases via \cbr and
applies reinforcement learning to adapt the solution to the situation at
hand.

Multiple contributions deal with CBR in the domain of injection molding.
A prototype recommendation system for parameter determination provides an
interface for manual parameter input and suggests corrections using
\cbr~\cite{KSL97}.
Another interactive user system for the shop floor determines parameters
first by \cbr and improves these through a rule-based system~\cite{SS97}.

Further research reviews different methods for parameter determination
injection molding~\cite{GZZL18}.
The authors identify \cbr as one of three main approaches.
They report no commercial or systematic solution since feedback on quality
remains challenging.
Instead of requiring manual reading and writing of parameters, a more
integrated approach incorporates the injection machine into the
system~\cite{ZZF07}.
However, it employs a very rudimental read/write approach without the goal
of creating a digital machine representative.
A concept of \dts in injection molding identifies all different phases of
the process and their linking~\cite{LLR18}.
However, it provides no defined method for the individual steps.
Our system focuses on the manufacturing procedure on the machine itself and
employs \cbr for this.

\section{Conclusion}
\label{sec:Conclusion}

To leverage \cbr over domain expertise into self-adaptive manufacturing, we
devised a modeling framework comprising multiple interrelated modeling
languages and integrate it into our architecture for
\dts~\cite{BDH+20,KMR+20,dalibor2020towards}.
We have presented a collection of modeling languages to support domain
experts in encoding their knowledge into \dts that perform self-adaptation at runtime.
This enables the \dts to react to unforeseen situations quickly and learn from past situations.
Models of these languages describe domain-specific cases and their
similarity and are processed by a modular \dt architecture that manages the
\cbr cycle of retrieving cases similar to the current situation, reusing
these to handle the situation, revising these if necessary, and retaining
these if the revisions were successful.
The realized framework is not tailored to one specific \cpps but can be customized to any other \cpps. 
The model-driven development simplifies developing \dts and can lead to more efficient
manufacturing, less misproduced goods, and, ultimately, reduced production cost.




\bibliographystyle{IEEEtran}

\bibliography{src/bib/sselit,src/bib/TB,src/bib/GB,src/bib/md}
  	
\end{document}